\documentclass{elsart1p}
\usepackage{graphicx}
\usepackage{amssymb}
\begin{document}
\begin{frontmatter}
\title{Photon multiplicity measurements at forward rapidity in the ALICE experiment at CERN}
\author{S.~K.~Prasad for the ALICE Collaboration}
\address{Variable Energy Cyclotron Centre, 1/AF Bidhan Nagar, Kolkata, INDIA.}
\begin{abstract}
We present the first preliminary results on photon
multiplicity measurements at LHC at forward pseudorapidity
(2.3 $\textless$ $\eta$ 
$\textless$ 3.9) in proton-proton
($pp$) collisions at $\sqrt{s}$ = 7 TeV.
The multiplicity distribution is found to be reasonably well explained by
a double Negative Binomial Distribution (NBD). The average photon multiplicity
increases logarithmically with $\sqrt{s}$.  It is found that none
of the models we used (PYTHIA6D6T, PHOJET, and HERWIG) could explain the data.
\end{abstract}
\end{frontmatter}
\section{Introduction}
One of the aims of studying the relativistic heavy-ion collisions is
to create, study and understand a system of deconfined states of quarks and
gluons commonly known as the {\it Quark-Gluon-Plasma} (QGP). A set of
proposed signatures provides evidences for the formation
of such a system in these collisions. Measurements in $pp$ collisions
define a baseline for drawing a definite conclusion in heavy-ion collisions.
Recently, measurements of multiplicity and
pseudorapidity distributions have been reported for charged particles
at midrapidity in $pp$ collisions for $\sqrt{s}$ = 0.9,
2.36, and 7 TeV by ALICE, CMS and ATLAS experiments at
CERN-LHC~\cite{alicemult2_4, ppmultlhc}. 
Photon measurements provide complementary information. The majority of
photons emitted from the reaction are decay products of produced
particles.
The multiplicity measurements at forward pseudorapidity 
represent an additional tool for testing the limiting fragmentation (LF)
hypothesis which is found to hold at lower energies~\cite{limitingfrag}.

In this article, we report the first measurement of the multiplicity
distribution of inclusive photons at forward pseudorapidities of 2.3
to 3.9 using the ALICE detector at the CERN-LHC for $pp$ collisions at
$\sqrt{s}$ = 7 TeV.

\section{The experiment and data analysis}
The present analysis uses data from the Photon Multiplicity Detector
(PMD). The PMD is a preshower detector with a three
radiation length thick lead converter sandwiched between two planes
of highly granular gas proportional counters. The information from one
of the detector planes, placed in front of the converter is used to
veto the charged particles, whereas the preshower data from the other
detector plane is used for photon
identification~\cite{pmdtdratdrnim}. In the present analysis data from
preshower plane only is used.

We use triggered event sample (OR between
the signals from the Silicon Pixel Detector and the VZERO, $\rm
MB_{OR}$ $\it INEL$~\cite{alicemult2_4}) which
corresponds to the minimum bias inelastic ({\it INEL}) events. Events
with vertex Z-position within
$\pm$ 10 cm from the
interaction point are analyzed. 
The total number of events used in the analysis after the
trigger and the vertex selections is around 390 K.
\section{Photon reconstruction and unfolding method}
The reconstruction of photons is achieved by finding
clusters of hits and then discriminating photons and hadrons.
A simple clustering algorithm with a contiguous cell search is employed
to find the clusters. The properties of each cluster with regard to
the number of cells (cluster $\rm N_{cell}$) contained in it and the total energy
deposition (ADC) have been obtained. 
Discrimination between photons and
hadrons is made by using a threshold method by putting a cut-off
in the cluster $\rm N_{cell}$ and ADC content.
\begin{figure}[here]
  \centering
  \includegraphics[height=4.7cm]{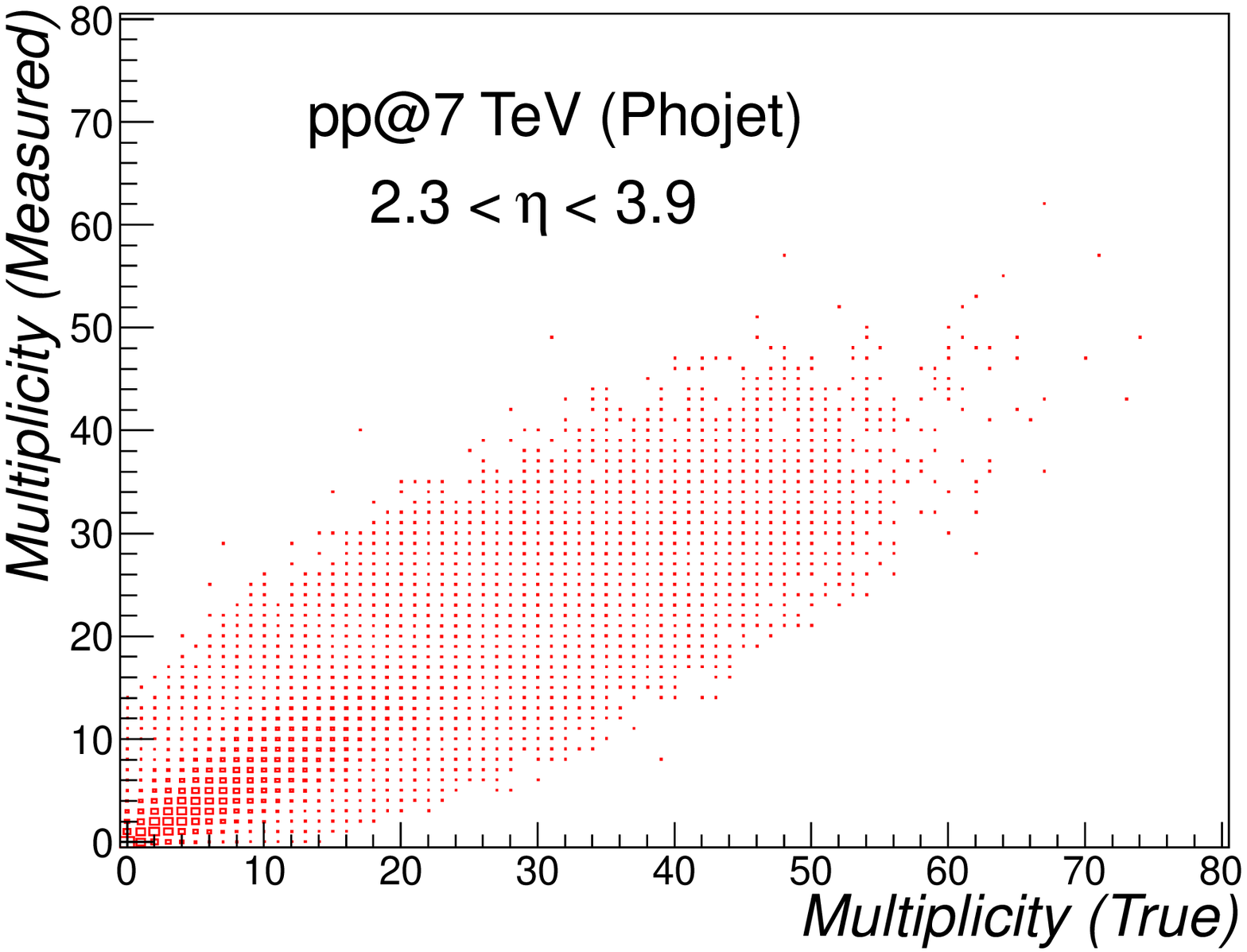}
  \includegraphics[height=4.7cm]{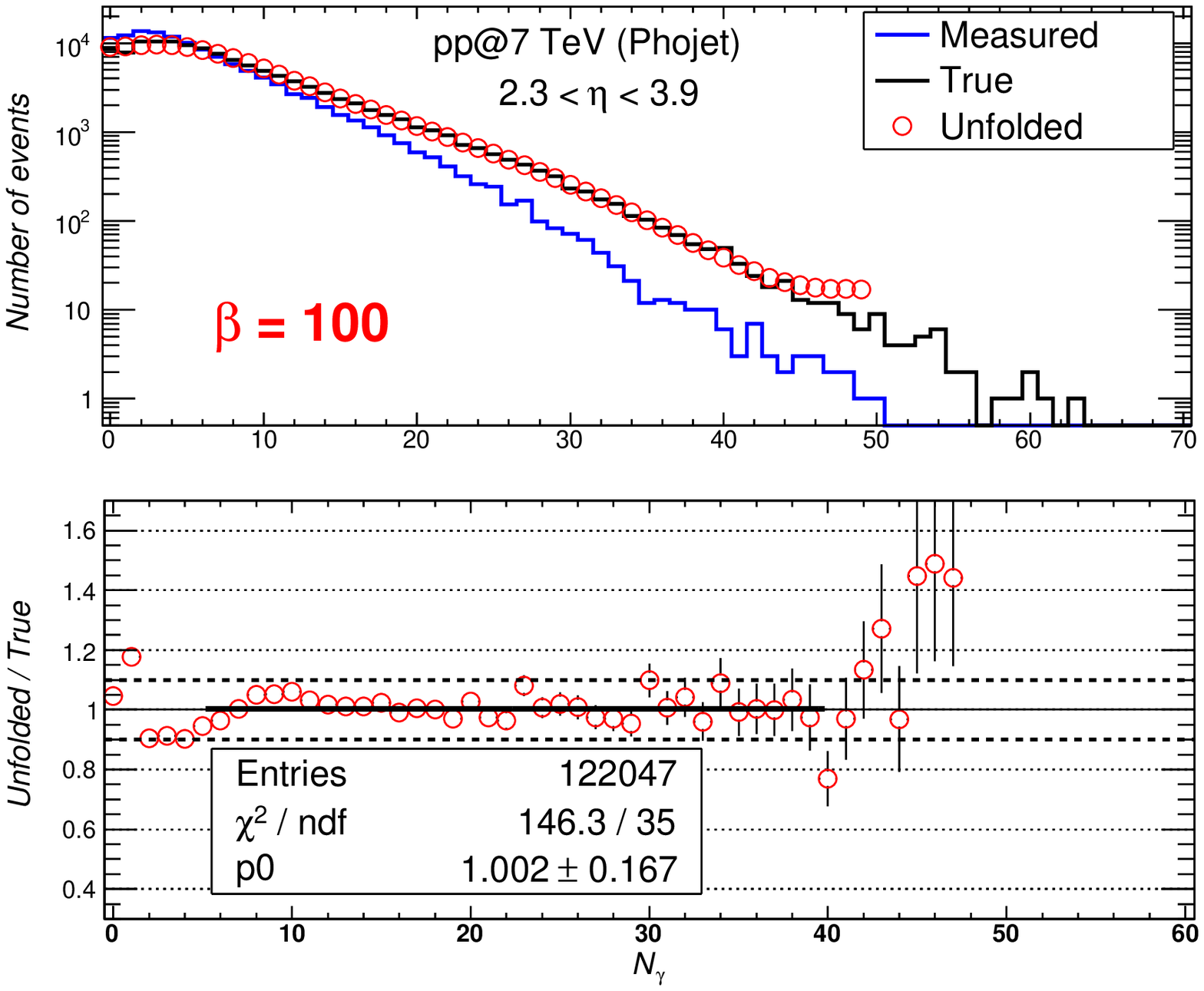}
  \caption{{\small (Color online) (Left) The detector response in terms of measured
    and incident photon multiplicity. (Right-Top) The distribution of true,
    measured, and unfolded multiplicity of photons. (Right-Bottom) The ratio
    unfolded/true. Results are obtained using PHOJET for $pp$ collisions at $\sqrt{s}$
    = 7 TeV for $\beta$ = 100.}}
\label{responsematrix_4}
\end{figure}

The method used to correct the measured raw distribution for
efficiency, acceptance and other detector effects, is based on unfolding 
with $\chi^2$-minimization with regularization, similar to what is
followed for charged particles measurement~\cite{alicemult2_4}.
In this method all the detector effects are described by a response
matrix, {\bf A}. The matrix element {\bf $A_{mt}$} gives
the conditional probability that an event with true multiplicity {\bf
  t} is measured as an event with the multiplicity {\bf m}
(see Ref.~\cite{thesissid} and references therein for detail). 
Figure~\ref{responsematrix_4}(Left) shows the PMD detector response in
terms of reconstructed photon multiplicity vs. true photon multiplicity
within 2.3 $\textless$ $\eta$ $\textless$ 3.9 in $pp$
collisions at $\sqrt{s}$ = 7 TeV using 1.2 M events from the PHOJET event
generator. The detector simulation is done using GEANT3 package in
ALICE environment. This 2-dimensional distribution forms the response
matrix to unfold the measured distribution in data.
The unfolded spectrum is found by minimizing
\begin{displaymath}
  \chi^2(U) = \sum_{m}{}\left ( \frac{g_m - \sum_{t}{}A_{mt}U_t}{e_m} \right)^2 + \beta P(U).
\end{displaymath}
where {\it $e_m$} is the error in the measurement {\it {\bf g}}, and {\it{\bf U}} is the
guessed spectrum. A regularization term $\beta${\it P(U)} is added to the
$\chi^2$-function to get rid of the spurious oscillations in the solution.
The coefficient $\beta$ is the weight factor.
Figure~\ref{responsematrix_4}(Right) shows the true, measured and unfolded
multiplicity distributions of photons (Top) and the ratio of unfolded to 
true distributions (Bottom) using simulated data. The ratio between
the unfolded and the true distribution is unity within $\pm$ 10\% up
to a multiplicity $N_\gamma$ $\textless$ $\sim$ 
40 indicating that the performance of the method is satisfactory.
\section{Results and Discussion}
\begin{figure}[here]
\centering
\includegraphics[height=4.6cm]{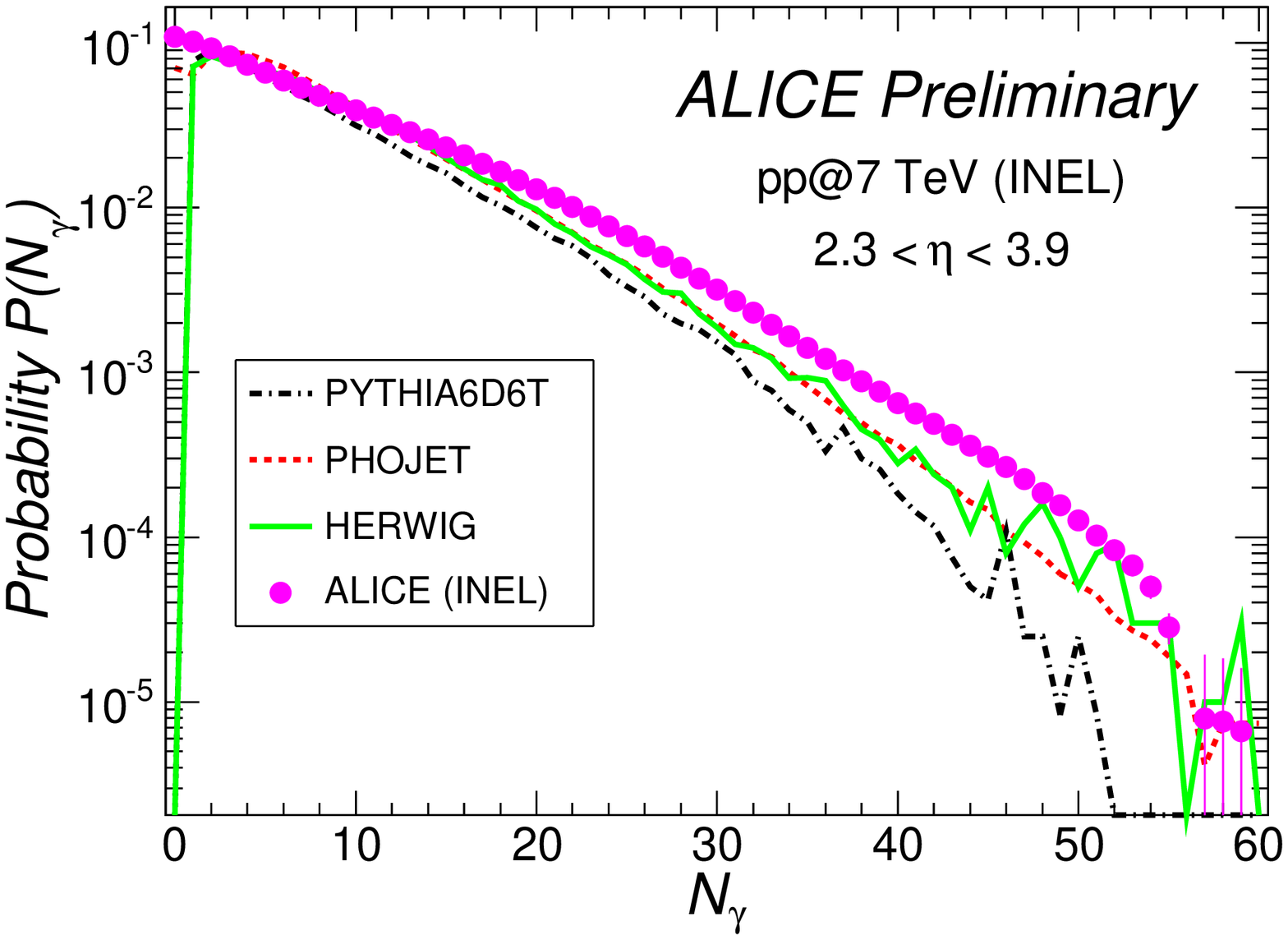}
\includegraphics[height=4.6cm]{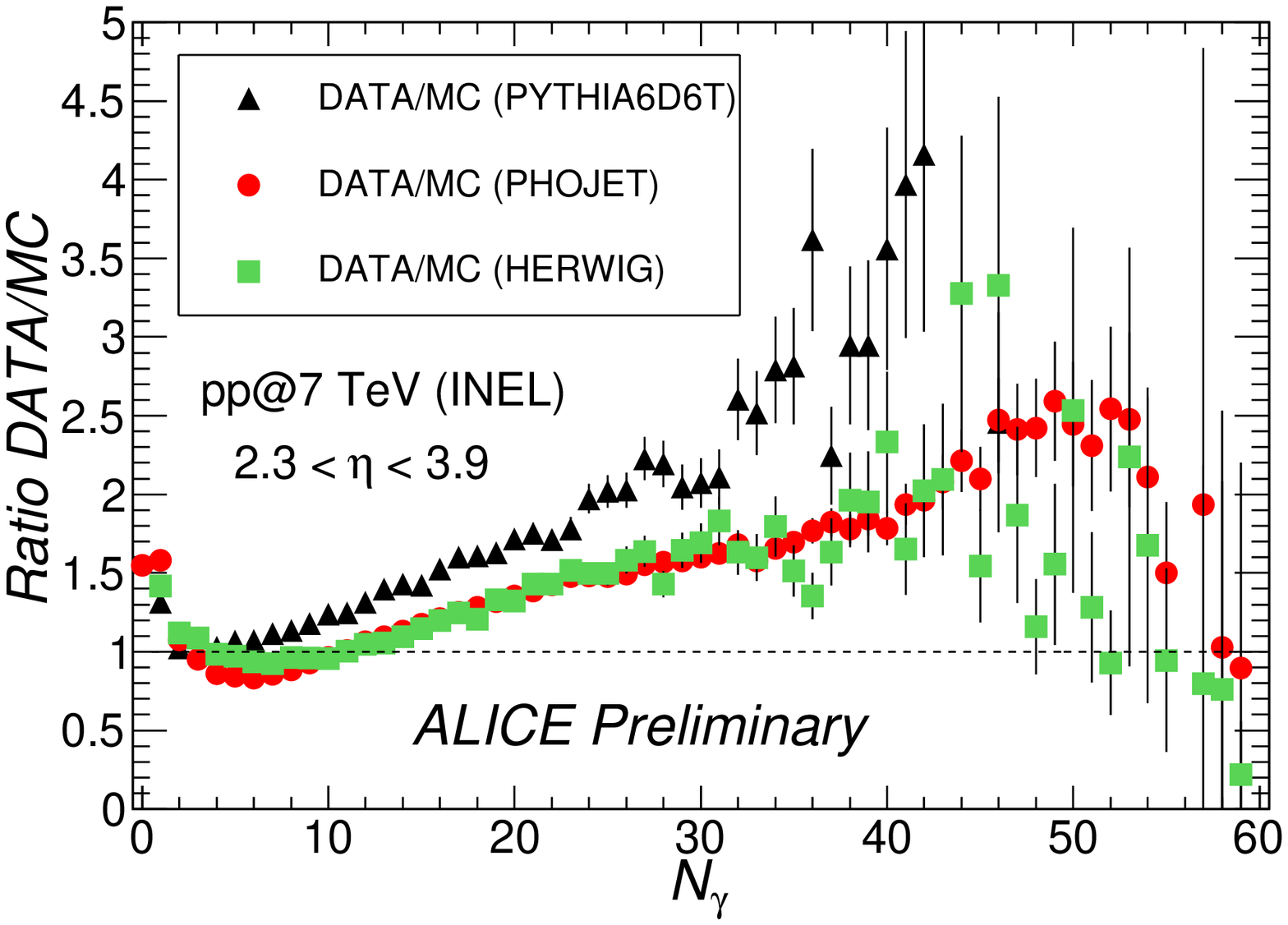}
\caption{{\small (Color online) (Left)
    $\rm N_{\gamma}$ distributions for 2.3
    $\textless$ $\eta$ $\textless$ 
    3.9, for {\it INEL} events at $\sqrt{s}$ = 7 TeV (Solid magenta
    circles). The lines are the predictions from PYTHIA6D6T
    (dashed black), PHOJET (dotted red), and HERWIG
    (solid green). (Right) The ratio of the multiplicities between the data
    and the model predications at 7 TeV
    (PYTHIA6D6T (black triangles), PHOJET (red circles), and HERWIG
    (green squares)). The error
    bars for data points represent statistical uncertainties in both
    the figures.}}
\label{mult7tevmc_4}
\end{figure}

\begin{figure}[here]
 \centering
 \includegraphics[height=4.3cm]{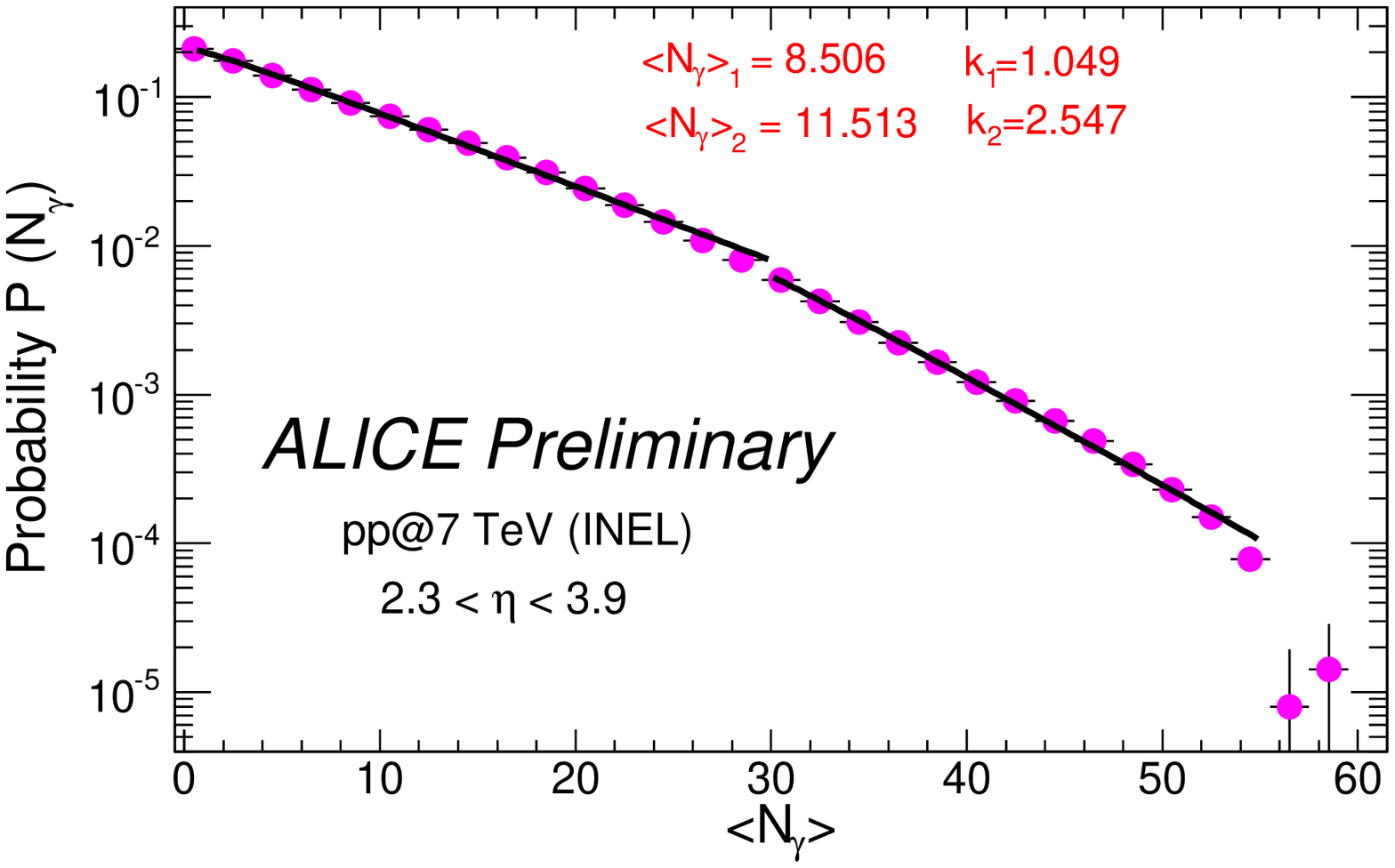}
 \includegraphics[height=4.3cm]{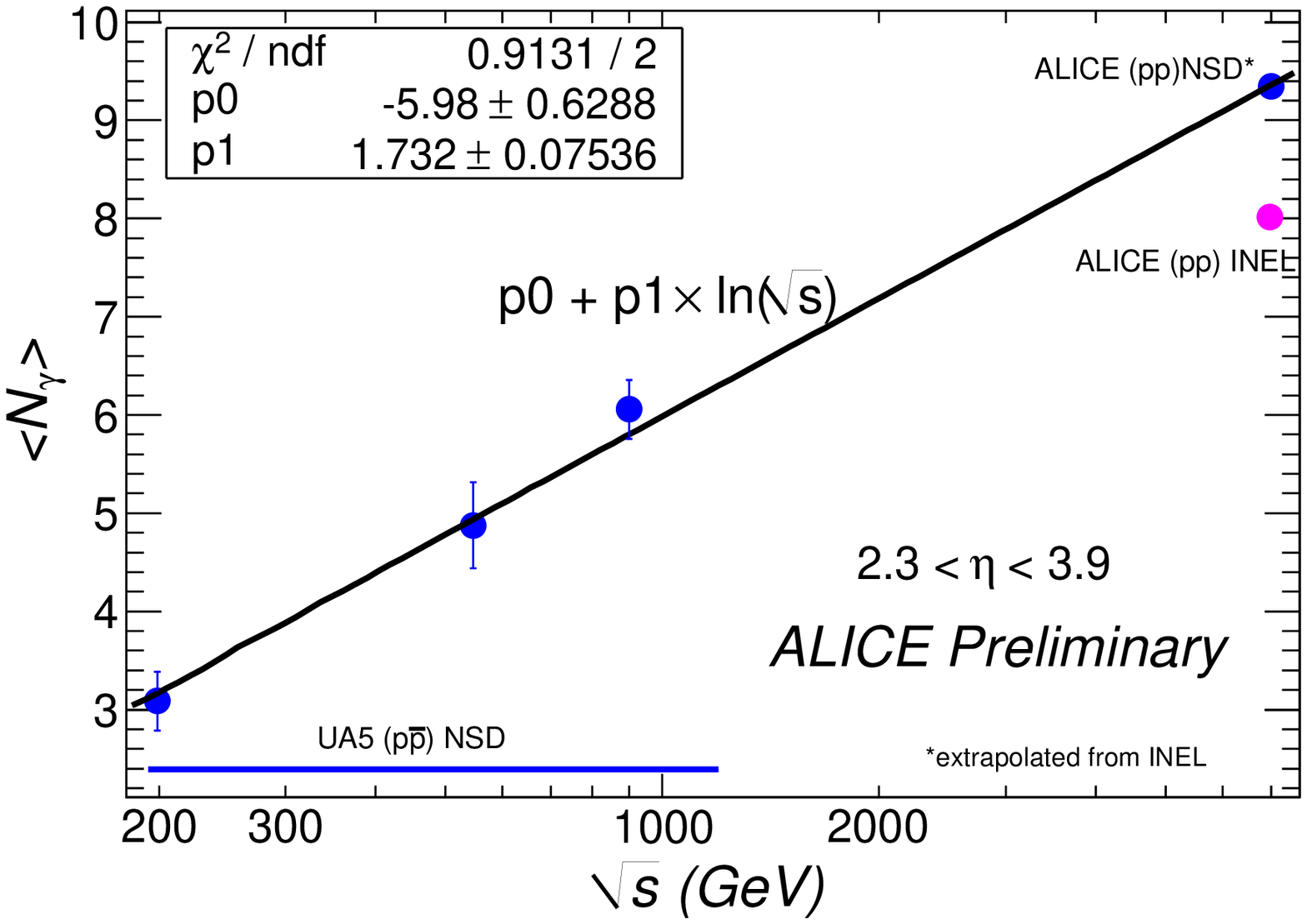}
\caption{{\small (Color online) (Left)
     $N_{\gamma}$ distributions for 2.3 $\textless$ $\eta$
     $\textless$ 3.9 for {\it INEL} $pp$ events at
     $\sqrt{s}$ = 7 TeV. The solid line shows double NBD fit. The error
     bars for data points represent statistical uncertainties.
     (Right) The average number of photons measured
     within 2.3 $\textless$ $\eta$ $\textless$ 3.9, for
     {\it NSD} events as a function of $\sqrt{s}$. The solid line is
     a logarithmic fit to the data. 
   }}
 \label{multnbdfit_4}
\end{figure}

Figure~\ref{mult7tevmc_4} (Left) shows the multiplicity distribution
of photons within 2.3 $\textless$ $\eta$ $\textless$ 3.9 compared
to the model predictions from PYTHIA6D6T, PHOJET, and HERWIG.
The ratios between the data and  
the models are shown in Figure~\ref{mult7tevmc_4} (Right). 
It is found that none of the models
reproduces the data. At higher multiplicities ($N_{\gamma}$ $\textgreater$ 10)
all models under-predict the data. However, at lower multiplicities 
($N_{\gamma}$ $\textless$ 10), PYTHIA under-predicts the data whereas 
PHOJET and HERWIG seems to over-predict the data.  
The multiplicity distribution of photons can be described by a double NBD
as shown in Figure~\ref{multnbdfit_4} (Left).
Studies show that a single NBD
explains the data up to a multiplicity of
$N_{\gamma} \sim$ 30 only, at higher multiplicities ($N_{\gamma}$
$\textgreater$ 30) it over-predicts the data.
Figure~\ref{multnbdfit_4} (Right) shows the $\sqrt{s}$ dependence
of the average number of photons per event within 2.3 $\textless$
$\eta$ $\textless$ 3.9  for non-single diffractive
({\it NSD}) events. 
The data points at lower energies are obtained from
Ref.~\cite{photonat200and900_4}. In Ref.~\cite{photonat200and900_4}
the UA5 experiment presented the results for {\it NSD} events, therefore, the
data points at 7 TeV, which are measured for {\it INEL} events,
are scaled to the corresponding {\it NSD} using PYTHIA simulations. The solid line is the fit to
the data points with a logarithmic function of the form p0 + p1
$\times$ $\ln{\sqrt{s}}$. 
The average number of photons within 2.3 $\textless$ $\eta$
$\textless$ 3.9 for {\it NSD} events increases with
$\sqrt{s}$, and the dependence is found to be logarithmic of the form
as mentioned earlier. For comparison our results for {\it INEL} are
also shown in Figure~\ref{multnbdfit_4} (Right).  
\section{Summary}
In summary, we have presented the first measurement of
multiplicity distribution of photons produced in $pp$ 
collisions at the LHC, at forward pseudorapidity (2.3 $\textless$
$\eta$ $\textless$ 3.9) using the PMD data at $\sqrt{s}$ = 7
TeV for INEL events.
The multiplicity distribution is found to be reasonably well explained by a
double NBD. The average photon
multiplicity for {\it NSD} events increases with $\sqrt{s}$ and the dependence
on $\sqrt{s}$ is found to be logarithmic of the form p0 + p1 $\times$
$\ln{\sqrt{s}}$.
The multiplicity distribution is compared with the available
model predictions from PYTHIA6D6T, PHOJET and HERWIG and it is found
that none of them could explain the data.

\end{document}